\documentstyle[preprint,aps,prbbib]{revtex}
\begin{document}
\def\gsim{\stackrel{>}{\sim}}
\def\lsim{\stackrel{<}{\sim}}
\def\cl{\centerline}
\def\half{ {1 \over 2} }
\def\J{ J }
\def\jl{ \J_{\sss L} }
\def\jr{ \J_{\sss R} }
\def\Q{  Q }
\def\O{ {\bf O} }
\def\ea                               {
     \epsilon_1                           }
\def\e                               {
     \epsilon                           }
\def\elr                               {
     \epsilon_{\sss L(R)}                           }
\def\el                               {
     \epsilon_{\sss L}                           }
\def\er                               {
     \epsilon_{\sss R}                           }
\def\ezero                               {
     \e_0                            }
\def\epeak                               {
     \e_{\rm peak}                            }
\def\efwhm                               {
     \e_{{\sss {\rm FWHM}}}                            }
\def\fl {f_{\sss L}(\e)}
\def\fr {f_{\sss R}(\e)}
\def\flw {f_{\sss L}(\w)}
\def\frw {f_{\sss R}(\w)}
\def\flr {f_{\sss L(R)}}
\def\mul {\mu_{\sss L}}
\def\mur {\mu_{\sss R}}
\def\nl {N_{\sss L}}
\def\nr {N_{\sss R}}
\def\mulr {\mu_{\sss L/R}}
\def\sp                                {
     \hskip 0.02in                            }
\def\ss                               {
     \scriptstyle                       }
\def\sss                               {
     \scriptscriptstyle                       }
\def\raise                             {
     {\sss +}                            }
\def\k                               {
     k                            }
\def\w                               {
     \omega                             }
\def\c                               {
     {\bf c}                            }
\def\spin                            {
     \sigma                             }
\def\spinbar                         {
     {\bar \spin}                  }
\def\cspin                               {
     {\bf c}_{\spin}                   }
\def\cspindag                          {
     {\bf c}_{\spin}^{\raise}               }
\def\fspin                               {
     {\bf f}_{\spin}                   }
\def\fspindag                          {
     {\bf f}_{\spin}^{\raise}               }
\def\fspindagbar                          {
     {\bf f}_{\spinbar}^{\raise}               }
\def\b                                   {
     {\bf b}                   }
\def\bdag                                   {
     {\bf b}^{\!\raise}                   }
\def\btwodag                                   {
     {\bf b}_2^{\!\raise}                   }
\def\fspindag                          {
     {\bf f}_{\spin}^{\raise}               }
\def\cbar                                {
     {\bf c}_{\spinbar}                   }
\def\cbardag                          {
     {\bf c}_{\spinbar}^{\raise}               }
\def\ckspin                               {
     {\bf c}_{\k\spin}                   }
\def\ckspindag                            {
     {\bf c}_{\k\spin}^{\raise}              }
\def\Vkspin                               {
     V_{\k\spin}                   }
\def\Vkbar                            {
     V_{\k\spinbar}                   }
\def\espin                           {
     \epsilon_{\spin}                   }
\def\ebar                        {
     \epsilon_{\spinbar}                }
\def\ekspin                           {
     \epsilon_{\k\spin}                   }
\def\ekbar                        {
     \epsilon_{\k\spinbar}                }
\def\sumk                            {
     \sum_{\k{\sss \in L,R} }                     }
\def\sumspin                         {
     \sum_{\spin}               }
\def\sumkspin                         {
     \sum_{\spin;\k{\sss \in L,R} }               }
\def\sumlr                         {
     \sum_{\sss L,R }               }
\def\sumklorr                         {
     \sum_{\k {\sss \in L(R)} }               }
\def\nup                              {
     n_{\sss\uparrow}                         }
\def\ndown                            {
     n_{\sss\downarrow}                         }
\def\nbar                             {
     \langle n_{\spinbar} \rangle              }
\def\nspin                            {
     \langle n_{\spin} \rangle              }
\def\nspinprime                            {
     \langle n_{\spin'} \rangle              }
\def\Gt                               {
     G_{\spin}(t)                         }
\def\Grt {G_{\spin}^{\,r}(t)}
\def\Gkt                               {
     G_{\k\spin}(t)                         }
\def\Gl        {
     G_{\spin}^<                         }
\def\Gg        {
     G_{\spin}^>                         }
\def\Glg       {
     G_{\spin}^{\stackrel{\sss <}{\sss >}}  }
\def\Gtwo                              {
     G_{\spinbar \spin}(t)                         }
\def\Ge                               {
     G_{\spin}(\e)                         }
\def\Grw                               {
     G^{\,r}_{\spin}(\w)                         }
\def\Glessw        {
     G_{\spin}^<(\w)                         }
\def\Fw                               {
     F_{\spin}(\w)                         }
\def\fbar   {\bar f_{\spin}(\w)}
\def\sigzero                          {
     \Sigma_{{\sss 0}\spin}              }
\def\sigone                          {
     \Sigma_{{\sss 1}\spin}              }
\def\cond                             {
     \sigma                               }
\def\Gamalr                            {
     \Gamma^{\sss L(R)}_{\spin}               }
\def\Gamazl                            {
     \Gamma^{\sss L}_{{\sss 0}\spin}              }
\def\Gamazr                            {
     \Gamma^{\sss R}_{{\sss 0}\spin}              }
\def\Gamazlr                            {
     \Gamma^{\sss L(R)}_{{\sss 0}\spin}              }
\def\Gama                            {
     \Gamma_{\spin}               }
\def\Gamal                            {
     \Gamma^{\sss L}_{\spin}               }
\def\Gamar                            {
     \Gamma^{\sss R}_{\spin}               }
\def\Gamblr                            {
     \Gamma^{\sss L(R)}_{{\sss 1}\spin}               }
\def\Gambl                            {
     \Gamma^{\sss L}_{{\sss 1}\spin}               }
\def\Gambr                            {
     \Gamma^{\sss R}_{{\sss 1}\spin}               }
\def\lifet {{1\over{\tau_{\sss \spin 1}}}   }
\def\ffd                            {
     f_{\sss FD}                               }
\def\eup                                {
    \epsilon_{\sss \uparrow}             }
\def\edown                              {
    \epsilon_{\sss \downarrow}             }
\def\kbt                                {
    k_{\!\sss B}T                            }
\def\tk                                 {
    T_{\!\sss K}                            }
\def\deltae                             {
    \Delta \e                              }
\def\gcfe                               {
    H_0 - \mul\nl - \mur\nr              }
\def\dqone                              {
    \delta_{{\ss \Q},1}               }
\def\dqn                              {
    \delta_{{\ss \Q},n}               }
\def\rhosw                            {
    \rho_{\sigma}(\w)                }
\def\tc                                 {
    T_{C}                       }
\def\sc                                 {
    S_{C}(-\infty,-\infty)                       }
\def\Zone                               {
    Z_{{\ss \Q} = 1}                                   }
\def\Zzero                               {
    Z_{{\ss \Q} = 0}                                   }
\def\Oqone                               {
    \langle \O \rangle_{{\ss \Q} = 1}                  }
\def\Oil                                 {
    \langle \O \rangle_{\ss i\lambda}                  }
\def\Oilone                                 {
    \langle \O \rangle_{\ss i\lambda}^{(1)}                  }
\def\Gil                                 {
    G_{\spin, i\lambda}^{r (1)}                  }
\def\Hlr                                    {
    H_{\spin}^{\sss L(R)}                    }
\def\Klr                                    {
    K_{\spin}^{\sss L(R)}                    }
\def\intinf                                 {
    \int_{-\infty}^{\infty}                  }
\def\dl { D^{\sss <} }
\def\dg { D^{\sss >} }
\def\dlg {D^{\stackrel{\sss <}{\sss >}}  }
\def\dgl {D^{\stackrel{\sss >}{\sss <}}  }
\def\dr { D^r }
\def\da { D^a }
\def\gl { G_{f\spin}^{\sss <} }
\def\gg { G_{f\spin}^{\sss >} }
\def\glg {G_{f\spin}^{\stackrel{\sss <}{\sss >}}  }
\def\gr { G_{f\spin}^r }
\def\ga { G_{f\spin}^a }
\def\pil { \Pi^{\sss <} }
\def\pig { \Pi^{\sss >} }
\def\pilg {\Pi^{\stackrel{\sss <}{\sss >}}  }
\def\pir { \Pi^r }
\def\pia { \Pi^a }
\def\sigl { \Sigma_{f\spin}^{\sss <} }
\def\sigg { \Sigma_{f\spin}^{\sss >} }
\def\siglg {\Sigma_{f\spin}^{\stackrel{\sss <}{\sss >}}  }
\def\sigr { \Sigma_{f\spin}^r }
\def\siga { \Sigma_{f\spin}^a }
\def\gkg {g_{\k\spin}^{\sss >}  }
\def\gkl {g_{\k\spin}^{\sss <}  }
\def\gklg {g_{\k\spin}^{\stackrel{\sss <}{\sss >}}  }
\def\gkgl {g_{\k\spin}^{\stackrel{\sss >}{\sss <}}  }
\def\eoms{equations-of-motion }
\def\eomas{EOM }
\def\eoma{EOM}
\def\eom{equations-of-motion}
\def\nca{non-crossing approximation}
\def\ncaa{NCA}
\def\ncaas{NCA }
\def\ncas{non-crossing approximation }
\def\doss{density of states }
\def\dos{density of states}
\def\lts{lifetime }
\def\lt{lifetime}
\def\squote{}
\def\quote#1#2#3#4{\squote {#1,\ {\sl#2}\ {\bf#3}, #4}.\par}
\def\qquote#1#2#3#4{\squote {#1,\ {\sl#2}\ {\bf#3}, #4};}
\def\nquote#1#2#3#4{\squote {#1,\ {\sl#2}\ {\bf#3}, #4}}
\def\book#1#2#3{\squote { #1,\ in {\sl#2}, edited by #3}.\par}
\def\nbook#1#2#3{\squote { #1,\ in {\sl#2}, edited by #3}}
\def\bbook#1#2#3{\squote { #1,\ in {\sl#2}, edited by #3}.}
\def\trans#1#2#3{[ {\sl #1} {\bf #2},\ #3 ]}
\def\prl{Phys. Rev. Lett.}
\def\prb{Phys. Rev. B}
\def\pr{Phys. Rev.}

\def\rpoint                          {}
\def\lpoint                          {}
\hsize 15 truecm
\title{The Anderson Model out of Equilibrium:\\  Non-Crossing-Approximation
Approach to Transport through a Quantum Dot}
\author{\large Ned S. Wingreen}
\address{NEC Research Institute, 4 Independence Way, Princeton, NJ 08540}
\author{\large Yigal Meir}

\address{Department of Physics, University of California\\
Santa Barbara, CA 93106}
\maketitle
\begin{abstract}
{The infinite-$U$ Anderson model is applied to transport
through a quantum dot. The current and density
of states are obtained via the \nca\ for two spin-degenerate levels
weakly coupled to two leads. At low temperatures, the Kondo
peak in the equilibrium density of states
strongly enhances the linear-response conductance.
Application of a finite voltage bias reduces the conductance
and splits the peak in the density of states.
The split peaks,
one at each chemical potential, are suppressed in
amplitude by a finite dissipative lifetime. We estimate
this lifetime perturbatively as the time to transfer an electron
from the higher chemical potential lead to the lower chemical
potential one. At zero magnetic field, the clearest signatures
of the Kondo effect in
transport through a quantum dot are the broadening,
shift, and enhancement
of the linear-response conductance peaks at low temperatures, and
a peak in the nonlinear differential conductance around zero bias.}
\end{abstract}

\section{Introduction}

The Kondo effect has been a focus of condensed matter research for
many years. Its essence -- the crossover from weak to strong
coupling between a magnetic impurity and a conduction-electron sea as
temperature is lowered -- has inspired both theory and
experiment.  While most aspects of the problem are now
well understood, studies have traditionally been confined to {\it equilibrium}
properties.\cite{fulde} For the case of a magnetic atom embedded in a
bulk metal,
achieving nonequilibrium may be deterring, but it is not
in the case of ``artificial atoms".\cite{vanhouten}
In particular, we predict that a quantum dot weakly coupled to its
leads is a Kondo system
in which nonequilibrium can be routinely achieved.\cite{selman,mwltwo}
More generally, an impurity or defect level in a small
structure where the applied bias is dropped
over a mesoscopic length\cite{beasley,ralph}
will be a nonequilibrium Kondo system.

Anderson's model \cite{anderson} for a Kondo impurity -- a site with discrete,
interacting levels coupled to a band -- has already
been used successfully to describe experiments on
quantum dots.\cite{averin,mwlone,beenakker}
The discrete spectrum of a single dot has been
observed by transport\cite{reed,paul,leo} and capacitance\cite{ashoori}
spectroscopy, while the strong on-site Coulomb interaction is
recognized\cite{been1} as the origin of periodic conductance
oscillations.\cite{paul,leo,meirav}
However, it is only the high
temperature regime that has been explored experimentally,
while it is at low temperatures that the Kondo effect emerges.

Since the Anderson Hamiltonian describes the quantum dot,
at low temperatures the dot must behave as a Kondo impurity.
In fact, Glazman \& Raikh\cite{glazman} and Ng \& Lee\cite{ng} have
argued that at zero-temperature equilibrium
the Kondo resonance in the density of states of
spin-degenerate levels will
produce
perfect transparency
of a quantum dot symmetrically coupled to its leads. More precisely,
for all chemical potentials between $\ezero$ and $\ezero + U$, where
$\ezero$ is the bare-level energy and $U$ is the interaction energy
(Fig. \ref{fig:schematic}),
the dot will have the conductance of an open channel, $2e^2/h$.
This is to be contrasted with the situation at temperatures
larger than $\Gamma$, the elastic width of the levels,
where the conductance consists of two resonances, at $\ezero$ and at
 $\ezero + U$.
 Since the
chemical potential of a quantum dot can effectively be swept by
changing the voltage on a nearby gate, the
Kondo effect will have a striking experimental signature in
low-temperature transport through a quantum dot.

Until now, however, only qualitative predictions have been made for
experimental observation of the Kondo effect in transport through
a quantum dot. Specifically, raising the temperature above the
relevant Kondo temperature is predicted to suppress the peak
in the density of states,
and, consequently, reduce the conductance.\cite{glazman,ng}
As the Kondo temperature near the conductance peak at $\ezero$
depends exponentially on the chemical potential,
$T_K(\mu)\sim \exp[-\pi(\mu-\ezero)/\Gamma]$, one expects that the
Kondo effect will enhance the conductance mainly for
$\ezero<\mu\lesssim\ezero+ {\rm few}\  \Gamma$.
Accordingly, Ng and Lee\cite{ng}
predicted that the finite temperature conductance vs. gate voltage
will consist of pairs of asymmetric peaks,\cite{alt}
separated by the Coulomb-interaction
energy $U$. In this work we present, for the first time,
a quantitative calculation of the lineshape of these conductance
peaks, via the non-crossing approximation.\cite{bickers} We show
that at experimentally accessible temperatures the Kondo effect
will leave the conductance peaks symmetric. The
Kondo effect will manifest itself, instead, in the broadening
of the peaks, the enhancement of their
amplitude, and the shift in their positions towards each other (for
each spin-degenerate pair), as the temperature decreases.

Furthermore, as the leads
coupled to a quantum dot are easily biased to nonequilibrium,
new physical questions which were not relevant
to magnetic impurities can also raised. In particular, what
happens to the Kondo effect out of equilibrium?\cite{selman,mwltwo}
Since at equilibrium the Kondo peak in the density of states
occurs at the chemical potential, the presence in nonequilibrium
of {\it two} chemical potentials must have a dramatic effect.

In this paper we try to answer the question of what new
behavior is present in the Anderson model out of equilibrium,
and to make quantitative predictions for experiment. Generalizing
the \ncas to nonequilibrium, using the Keldysh
formalism, we find that
a voltage bias between the left and right leads
causes the Kondo peak in the density of states to split,
leaving a peak in the density of states at
the chemical potential of each lead
(Fig. \ref{fig:fdos}).
The amplitudes of
these split peaks are
suppressed by a finite nonequilibrium lifetime,
due to dissipative transitions
in which electrons are transferred from the higher chemical potential lead
to the lower chemical potential one.
The narrowness of the Kondo peak in the density of states, and its
splitting and suppression, lead to a rapid
decrease of conductance with increasing bias.
The resulting peak in the nonlinear differential conductance
is likely to be the most accessible experimental signature of
the Kondo effect in quantum dots.

We begin this paper with a general formulation of nonequilibrium transport
through an Anderson impurity in the limit of an infinite
on-site interaction energy, $U$. (section~\ref{general}).  Short discussions of
the mapping into a ``slave-boson'' Hamiltonian and of the
Keldysh formalism are presented.  The \ncas is then introduced
(section~\ref{noncross}) and the numerical methods outlined
(section~\ref{numerical}).
Results of the \ncas are presented for both equilibrium
and nonequilibrium transport (section~\ref{results}). The theoretical
interpretation of the results is discussed (section~\ref{theory}) as well
as the implications for experiment (section~\ref{experiment}).
An appendix is included to demonstrate the current conserving
property of the \nca.

\section{The Nonequilibrium Anderson Model}
\subsection{General Formulation}
\label{general}
\subsubsection{The Model}
We model the quantum dot and its leads by the Anderson
Hamiltonian\cite{anderson}
\begin{equation}
 H = \sumkspin\!\! \ekspin \ckspindag \ckspin
  + \sumspin \espin \cspindag \cspin
  + \half U \sumspin \sum_{\spin'\neq\spin} n_{\spin}n_{\spin'}
+ \sumkspin \!
       (\Vkspin \ckspindag \cspin + h.c.),\label{eq:H}
\end{equation}
where $\ckspindag (\ckspin)$  creates (destroys) an electron with momentum $k$
and spin $\spin$ in one of
the two leads, and $\cspindag (\cspin)$  creates (destroys) a
spin-$\spin$ electron on the quantum dot. The spin quantum number
$\spin$ may also represents orbital degeneracies as in the magnetic
impurity problem,\cite{anderson} though, experimentally, these
degeneracies are likely to be lifted by disorder in quantum dots.
In the following we will focus on spin-degenerate
states.
The third term describes the Coulomb interaction among electrons
on the dot. We assume that $U \rightarrow \infty$, forbidding
double occupancy. This is appropriate for quantum dots
where, typically, $U (\sim 1 {\rm meV})$ is a hundred times larger
than the coupling to the
leads,\cite{ethan} $\Gamma (\sim 10 \mu{\rm eV})$.
The fourth term describes the hopping between the leads and the dot,
and determines this coupling strength via
\begin{equation}
 \Gamalr(\w)  = 2\pi\! \sumklorr |\Vkspin|^2 \,
   \delta(\w - \ekspin). \label{eq:width0}
\end{equation}

Our aim is to calculate the current through the quantum dot, $J$,
which for the case of proportionate coupling to the leads,
$\Gamal(\w) = \alpha \Gamar(\w)$, can be
expressed\cite{selman,mwltwo,mwone} in
terms of the density of states, $\rhosw$, as
\begin{equation}
\J = {e\over \hbar} \sum_{\spin} \intinf\!\!
d\w \left[ \flw - \frw \right]
\,\Gama(\w)\, \rhosw,
 \label {eq:J}
\end{equation}
where
$\Gama(\w) = \Gamal(\w)\Gamar(\w)/\left[\Gamal(\w)+\Gamar(\w)\right]$.
The density of states is given by
\begin{equation}
 \rhosw = -{1\over\pi} {\rm Im}\, \Grw, \label{eq:dos}
\end{equation}
where $\Grw$ is the Fourier transform of the retarded Green function,
\begin{equation}
 \Grt = -i\theta(t) \langle \{ \cspin(t),\cspindag(0) \}
   \rangle. \label{eq:Gtime}
\end{equation}

\subsubsection{The Slave-Boson Hamiltonian}
Diagrammatic techniques are reliable when the expansion
parameter is
a small quantity. For an Anderson impurity with
$U \rightarrow \infty$,
it is natural
to perturb in the hopping strength. However, the standard diagrammatic
approach also requires that the unperturbed Hamiltonian be
noninteracting, {\it i.e.}, quadratic in the second-quantized
operators. In the limit of infinite
$U$, the bare Hamiltonian can be made quadratic by transforming
the Hamiltonian (\ref{eq:H}) into a new Hamiltonian, expressed
in terms of new local operators.\cite{barnes,coleman}
 These operators create the three
possible states of the site: a boson operator $\bdag$,
which creates an empty site, and two fermion operators,
$\fspindag$, which create the
singly occupied states.
The ordinary electron operators on the site, which transform the
empty site into a singly occupied site or vice versa, are decomposed
into a boson operator and a fermion operator,
\begin{eqnarray}
 \cspin(t) &= \bdag(t)\,\fspin(t)  \cr
 \cspindag(t) &= \fspindag(t)\,\b(t).
\label{eq:decomp}
\end{eqnarray}
The ``slave boson'' in (\ref{eq:decomp}) acts as a
bookkeeping device which prevents
double occupancy of the site:
when an electron creation operator
acts on an occupied site,
the boson part acting on the vacuum annihilates the state,
$\cspindag\, \fspindagbar\,|\Omega\rangle =
\fspindag\,\b\,\fspindagbar\,|\Omega\rangle = 0$. (In these expressions
$|\Omega\rangle $ is the vacuum state.)

In the ``slave-boson" representation, the Hamiltonian for the
infinite-$U$ Anderson model becomes
\begin{equation}
 H = \sumkspin\!\! \ekspin \ckspindag \ckspin
  + \sumspin \espin \fspindag \fspin
+ \sumkspin \!
       (\Vkspin \ckspindag \bdag  \fspin + h.c.).\label{eq:Hboson}
\end{equation}
The first two terms form the unperturbed, quadratic Hamiltonian and
the last term, which represents hopping between site and leads, can be
handled as a perturbation. The fermions and boson are treated as
ordinary particles in the perturbation expansion. For example,
the lowest order diagrams are shown for the
boson and fermion propagators in Fig. \ref{fig:pt}.
 While summation of a few low-order diagrams is possible,\cite{barnes,pt}
techniques are also available to sum whole classes of diagrams.
In the end, whatever approach is taken, properties of the physical
electrons can be constructed from the results for the boson and fermions.

There is, however, an added constraint,
as the site can only be in
one of the states $\bdag\,|\Omega\rangle$ and $\fspindag\,|\Omega\rangle$.
Accordingly, in all physical states,  the number of bosons plus the number
of fermions,
\begin{equation}
  \Q = \bdag\,\b + \sum_{\spin} \fspindag\,\fspin, \label{eq:Qdef}
\end{equation}
must be equal to unity. We will show below, when we describe the
Keldysh diagrammatic approach, how this constraint is dealt with.

\subsubsection{The Keldysh Formalism}

Previous diagrammatic calculations using the ``slave-boson"
representation\cite{bickers} have addressed equilibrium properties
of the Anderson model. Since our focus is on nonequilibrium
properties a different approach is required.
Specifically, we employ the Keldysh\cite{keldysh,caroli}
rather than the Matsubara\cite{matsubara}
formalism. The main
complication
with nonequilibrium is that the
basis of the equilibrium diagrammatic approach, the fact that
 the state
of the system at $t = +\infty$ is identical to the state of the
system at $t = -\infty$, up to a phase
(Gell-Mann and Low theorem \cite{gellmann}),
is no longer valid.
Since in a nonequilibrium
system real dissipation can occur, the state of the system is
in general not known at $t = +\infty$ and one must relate all
quantities to the state of the system at $t = -\infty$. In
practice, this means that instead of having integrals from
$t = -\infty$ to $t = +\infty$ as in the usual zero-temperature
formulation,\cite{zerot} all integrals have to be carried out along a path
that starts and ends at $t = -\infty$ (Fig. \ref{fig:contour}).
Consequently, a Green function will depend not only on the
times at which the operators act, but also on the corresponding
branch of the contour. Thus the Green functions carry additional
indices, and the usual perturbation expansion, or the Dyson
equation, takes a matrix form. In all, there are three independent
types of two-particle Green functions in nonequilibrium.
It is convenient to choose,
in addition to the retarded Green function (\ref{eq:Gtime}),
the two Green functions:
\begin{eqnarray}
 \Gl(t) &= i\langle \cspindag(0) \cspin(t)\rangle \cr
 \Gg(t) &= - i\langle \cspin(t) \cspindag(0)\rangle,
\label{eq:Gless}
\end{eqnarray}
as they carry information on the occupation of the site.

For the problem at hand, the starting point at $t = -\infty$ is
the Anderson impurity and the leads unconnected
and separately at equilibrium, possibly with
different chemical potentials. Formally, the hopping is turned on
slowly, and nonequilibrium properties are evaluated long after
the hopping is fully established, when a steady state has been
achieved,
but before current flow
has changed the chemical potentials deep in the leads.

Before applying the Keldysh formalism to the ``slave-boson"
Hamiltonian (\ref{eq:Hboson}), one has to overcome
the difficulty associated with the constraint that
the physical states are
restricted to the $\Q = 1$ ensemble. Specifically,
it is not convenient to perform diagrammatic calculations
(Keldysh or Matsubara) in a restricted ensemble since Dyson's
equation does not apply.
Instead, we introduce a complex chemical potential,\cite{bickers}
calculate diagrams in an unrestricted ensemble
treating the hopping as a perturbation,
and finally use the complex chemical potential to
project to the $\Q = 1$ subspace. In practice, this projection
corresponds to keeping only an easily identified subset of diagrams.

To formulate the Keldysh diagrammatic theory in terms of a complex
chemical potential, it is convenient to start with a formal
expression for expectation values in nonequilibrium.
In the $\Q=1$ ensemble,
the expectation value of an operator $\O$, can be written as
\begin{equation}
 \Oqone = {1 \over \Zone} {\rm Tr}\,
 \Bigl\{ e^{-\beta(\gcfe)}\, \dqone\, \tc [ \sc \O] \Bigr\},
  \label{eq:Oqdef}
\end{equation}
where $\tc$ orders operators along the Keldysh contour
(Fig. \ref{fig:contour}) and
the partition function is given by
\begin{equation}
 Z_{{\ss \Q} = n}  = {\rm Tr}\,
 \Bigl\{ e^{-\beta(\gcfe)}\, \dqn\, \tc [\sc] \Bigr\},
  \label{eq:Zdef}
\end{equation}
with $\Q = 1$.
In (\ref{eq:Oqdef}), the system evolves under the action
of the Hamiltonian so that
\begin{equation}
 \sc = \exp\Bigl[-i\oint_C dt' H(t')\Bigr]. \label{eq:Sdef}
\end{equation}
Importantly,
the operator $\O$ may include parts acting at different times,
{\it e.g.}, $\O = i\cspindag(0)\,\cspin(t)$ would give the
nonequilibrium expectation value of the ``lesser" Green function,
$\Gl(t)$.
Since the Hamiltonian commutes with the sum of bosons and fermions, $\Q$,
the projection to the $\Q =1$ ensemble is accomplished once and for all
by the factor $\dqone$ in (\ref{eq:Oqdef}). It is not necessary to
include a chemical potential for the impurity since
local expectation values in the coupled system
are independent of the initial state
of the impurity.

To transform to an ensemble where $\Q$ is unconstrained,\cite{bickers} one
rewrites the Kronecker delta as an integral over a complex chemical
potential, $i\lambda$,
\begin{equation}
 \dqone = {\beta \over {2 \pi}} \int_{-\pi/\beta}^{\pi/\beta}
  \!\!d\lambda\, e^{-i\beta\lambda(\Q - 1)}. \label{eq:qone}
\end{equation}
Dividing both numerator and denominator of (\ref{eq:Oqdef}) by
$\Zzero$, gives\cite{oiszero}
\begin{eqnarray}
 \Oqone =& &\displaystyle{ {\Zzero  \over \Zone} \,
  {\beta \over {2 \pi}} \int_{-\pi/\beta}^{\pi/\beta}
  \!\!d\lambda\, e^{i\beta\lambda} \Oil} \nonumber\\[10pt]
  =& &\displaystyle{ {\Zzero  \over \Zone}\, \Oilone},
  \label{eq:Oq}
\end{eqnarray}
where
\begin{equation}
 \Oil = {1 \over {Z_{i\lambda}} } {\rm Tr}\,
 \Bigl\{ e^{-\beta(\gcfe + i\lambda\Q)}\, \tc [ \sc \O] \Bigr\}.
  \label{eq:Oildef}
\end{equation}
In (\ref{eq:Oq}), $\Oilone$ is the coefficient of the term
of order $\exp(-i\beta\lambda)$ in $\Oil$.
The important point is that $\Oil$ in (\ref{eq:Oildef})
is in the standard form for diagrammatic perturbation
theory since the trace in (\ref{eq:Oildef}) and in
the partition function,
\begin{equation}
 Z_{i\lambda} = {\rm Tr}\,
 \Bigl\{ e^{-\beta(\gcfe + i\lambda\Q)}\, \tc [ \sc] \Bigr\},
  \label{eq:Zildef}
\end{equation}
are taken over all states without restriction to $\Q=1$.

According to (\ref{eq:Oq}),
the nonequilibrium expectation value of an operator, $\O$,
in the $\Q=1$ ensemble
has two contributions: a normalization factor
$\Zzero/\Zone$, and the coefficient of $\exp(-i\beta\lambda)$
for the same operator in the $i\lambda$ ensemble. The normalization
can be obtained from the identity $\langle \Q \rangle_{{\sss Q} = 1}
= 1$, which implies
\begin{equation}
 {\Zone \over \Zzero} = \langle \bdag\, \b\rangle_{i\lambda}^{(1)}
  +  \sum_{\spin} \langle \fspindag\, \fspin \rangle_{i\lambda}^{(1)}.
  \label{eq:Znorm}
\end{equation}
The expectation value $\Oilone$, as well as the expectation values
appearing on the right-hand side of Eq. (\ref{eq:Znorm}) can be obtained
diagrammatically.

\subsection{Non-crossing Approximation}
\label{noncross}

To obtain a well-behaved density of states from the
nonequilibrium perturbation
theory, one needs some way of summing
diagrams to all orders in the hopping. In finite-order perturbation
theory there are divergences associated with the bare-levels,
$\e_{\spin}$,
and, at $T=0$, logarithmic divergences near the chemical potentials
due to the
Kondo effect.\cite{pt}
To control these divergences, we employ the \nca, which
has been used successfully to treat the infinite-$U$
Anderson model in equilibrium.\cite{bickers} As can be seen from
Fig. \ref{fig:pt}, at lowest order in perturbation theory
the boson self-energy involves the fermion propagator
while the fermion self-energy involves the boson propagator. By
using the two relations
self-consistently -- the \nca, see Fig. \ref{fig:nca} -- one
obtains a set of coupled integral equations, which can be
solved numerically.
Solving these self-consistent equations
corresponds to summing a subset of diagrams to all orders in
the hopping matrix element $V$. It can be shown\cite{bickers} that
all diagrams of leading order
in $1/N$,
where $N$ is the number of spin degrees of freedom,
are included in this subset.
Therefore, the non-crossing approximation is expected to be a
quantitative approach in the limit of large $N$. Importantly,
comparison to exact results\cite{cox} has shown that the \ncas is also
quantitative for the
case $N=2$, of interest for quantum dots.
Here, we generalize the \ncas to nonequilibrium.  The equations will
involve not only the retarded Green function, but also the lesser
and greater ones, leading to slightly more complicated
equations than at equilibrium.

Since our goal is to calculate the nonequilibrium current,
we will calculate first the density of states for the Anderson
impurity, $\rhosw$, and obtain the current from Eq. (\ref{eq:J}).
To find the density of states for $U \rightarrow \infty$, we
need the retarded Green function (\ref{eq:dos}) in the ensemble
with complex chemical potential (\ref{eq:Oq}),
\begin{equation}
 \rhosw = {\Zzero \over \Zone}\, \Bigl[ -{1 \over \pi}
 {\rm Im}\, \Gil(\w) \Bigr].
  \label{eq:rhosw1}
\end{equation}
Within the \nca, the retarded Green function
is expressed in
terms of the full propagators for boson and fermion as
\begin{eqnarray}
 \Gil(t) \equiv\ & -i\theta(t) \langle \{ \cspin(t),\cspindag(0) \}
   \rangle_{i\lambda}^{(1)} \cr
   \stackrel{\rm {\sss NCA}}{=}& -i\theta(t)\, \Bigl
          [\dg(-t)\,\gl(t) - \dl(-t)\,\gg(t)\Bigr],
\label{eq:Gilt}
\end{eqnarray}
where
\begin{eqnarray}
 \dg(t) \equiv& -i\langle\b(t)\,\bdag(0)\rangle_{i\lambda}^{(0)} \cr
 \dl(t) \equiv& -i\langle\bdag(0)\,\b(t)\rangle_{i\lambda}^{(1)} \cr
 \gg(t) \equiv&  -i\langle\fspin(t)\,\fspindag(0)\rangle_{i\lambda}^{(0)} \cr
 \gl(t) \equiv& \ \ \  i\langle\fspindag(0)\,\fspin(t)\rangle_{i\lambda}^{(1)}.
\label{eq:Gils}
\end{eqnarray}
Equation (\ref{eq:Gilt}) is straightforward to obtain by decomposing
the electron operators into boson and fermion operators
(\ref{eq:decomp}) and then factorizing the boson and fermion parts.
The latter step corresponds to a neglect of vertex corrections.
Since each term in (\ref{eq:Gilt}) contains exactly one ``lesser"
operator, with a boson or fermion lowering operator acting directly
to the right, the overall result is $O(\exp(-i\beta\lambda))$ as
required.

Because the Hamiltonian is time independent,
it is simplest to evaluate the boson and fermion Green functions
in the frequency representation.
The physical density of states is then given by
\begin{equation}
 \rhosw =  {1 \over {4\pi^2}}\, {\Zzero \over \Zone}
 \intinf\!\! d\w'\, \Bigl[ \dg(\w')\,\gl(\w + \w')
                              - \dl(\w')\,\gg(\w + \w') \Bigr].
  \label{eq:dostwo}
\end{equation}
The \ncas is represented diagrammatically in Fig. \ref{fig:nca}:
the boson and fermion propagators are each assigned a single
self-energy bubble (albeit determined self-consistently) and
the self-energies are iterated to all orders via Dyson's equation.
Standard manipulation of the nonequilibrium Dyson equations
then leads to\cite{mahan}
\begin{eqnarray}
 \dlg(\w) =& \dr(\w)\, \pilg(\w)\, \da(\w) \cr
 \glg(\w) =& \gr(\w)\, \siglg(\w)\, \ga(\w),
  \label{eq:Glgs}
\end{eqnarray}
where, the self-energies are given by
\begin{eqnarray}
 \pilg(\w)  =& \displaystyle{-{i \over {2\pi}} \intinf\!\!\!
                        d\w'\!\!\! \sumkspin
  \!|\Vkspin|^2 \, \gkgl(\w' - \w)\, \glg(\w')} \nonumber\\[10pt]
 \siglg(\w)  =& \displaystyle{ {i \over {2\pi}} \intinf\!\! d\w'\!\! \sumk
  \!|\Vkspin|^2 \, \gklg(\w - \w')\, \dlg(\w')}.
  \label{eq:selfs}
\end{eqnarray}
In (\ref{eq:selfs}), the small $g$'s are the Green functions of
electrons in the leads not coupled to the site,\cite{baregs}
\begin{eqnarray}
 \gkg(\w)  =& -2\pi i\, [1 - \flr(\w)]\, \delta(\w - \ekspin) \cr
 \gkl(\w)  =& \ 2\pi i\, \flr(\w)\, \delta(\w - \ekspin).
  \label{eq:gleads}
\end{eqnarray}

Several other relations are required to close the equations for the
\nca. The retarded Green functions for the boson and fermions
in (\ref{eq:Glgs}) are given by\cite{mahan}
\begin{eqnarray}
 \dr(\w)  =& \displaystyle{ {1 \over {w - \pir(\w)} } }\nonumber \\[10pt]
 \gr(\w)  =& \displaystyle{ {1 \over {w - \espin - \sigr(\w)} } },
  \label{eq:grs}
\end{eqnarray}
where the retarded self-energies are Hilbert transforms of the greater
self-energies
\begin{eqnarray}
 \pir(\w)   =& \displaystyle{ {i \over {2\pi}} \intinf \!\!d\w' \,
       { {\pig(\w')} \over {\w - \w' +i\eta} }} \nonumber\\[10pt]
 \sigr(\w)  =& \displaystyle{ {i \over {2\pi}} \intinf \!\!d\w' \,
       { {\sigg(\w')} \over {\w - \w' +i\eta} }}.
  \label{eq:selfrs}
\end{eqnarray}
The advanced Green functions $\da$ and $\ga$ in (\ref{eq:Glgs}) are complex
conjugates of the retarded Green functions $\dr$ and $\gr$.
Eqs. (\ref{eq:selfrs})
follow because, by definition, all retarded Green functions and
self-energies can be written as a difference of greater and
lesser functions, $G^r(t) = \theta(t)[G^{\sss >}(t) - G^{\sss <}(t)]$.
In the $i\lambda$ ensemble, the lesser functions for the
boson and fermions are
$O[\exp(-i\beta\lambda)]$ and must be dropped from the retarded
functions which are  $O(1)$.
One therefore has the useful relations:
\begin{eqnarray}
 \dg(\w)  =& 2i\, {\rm Im}\,  \dr(\w) \cr
 \gg(\w)  =& 2i\, {\rm Im}\,  \gr(\w),
  \label{eq:ggs}
\end{eqnarray}
for the boson and
fermion Green functions, and
\begin{eqnarray}
 \pig(\w)   =& 2i\, {\rm Im}\, \pir(\w) \cr
 \sigg(\w)  =& 2i\, {\rm Im}\, \sigr(\w),
  \label{eq:selfgs}
\end{eqnarray}
for the self-energies.

The closed set of equations for the \ncas can be solved iteratively.
In practice, we start with an initial guess for the greater boson
Green function, $\dg(\w)$, calculate $\sigr(\w)$ for each spin by
combining (\ref{eq:selfs}) and (\ref{eq:selfrs}), and use
(\ref{eq:grs}) to get $\gr(\w)$.
The values for the greater fermion Green functions, from
$\gg(\w) = 2i{\rm Im}\,\gr(\w)$,
can then be used in a parallel way to obtain an
improved $\dg(\w)$. This procedure is iterated to convergence.
A similar procedure is then followed for the
lesser Green functions. Following an initial guess for $\dl(\w)$,
the fermion self-energies, $\sigl(\w)$ are obtained from
(\ref{eq:selfs}), and $\gl(\w)$ is determined from (\ref{eq:Glgs}).
The steps are repeated for $\dl(\w)$, and the process iterated
to convergence. Finally, the physical density of states, $\rhosw$
is evaluated by the convolution of the boson and fermion Green
functions in (\ref{eq:dostwo}).

\subsection{Numerical Methods}
\label{numerical}

In this section, we describe in greater detail
the numerical procedures we have
used to solve the nonequilibrium, self-consistent equations for
the \nca.
Following the equilibrium work,\cite{bickers,cox}
we take the energy dependence of the
coupling between the site and the leads (\ref{eq:width0})
to be Lorentzian
\begin{equation}
 \Gamalr(\w)  = 2\pi\! \sumklorr |\Vkspin|^2 \,
   \delta(\w - \ekspin)
  \equiv  \Gamazlr { {W^2} \over { (\w - \elr)^2 + W^2}}.
 \label{eq:width1}
\end{equation}
The finite width, $W$, reflects the finite bandwidth in the leads
and is necessary to prevent ultraviolet divergence of the
results.\cite{uv} In principle, the bands in the leads can be
centered at different energies, but the
validity of Eq. (\ref{eq:J}) for the current
requires $\Gamal(\w) = \alpha \Gamar(\w)$, so we
take $\el = \er = \ezero = 0$, throughout.
[An expression (\ref{eq:currentlr})
for the current in the absence
of this condition is given in the appendix.]
The choice of a Lorentzian form allows
a simplification of the self-consistent equations.\cite{coxthesis}
In general, to iterate the \ncas equations, the retarded
self-energies for the boson and fermions (\ref{eq:selfrs})
must be evaluated by double integrals over the greater
Green function of the other species.
For the the Lorentzian coupling, however, one of these integrals can
be performed analytically. First, combining Eqs. (\ref{eq:selfs})
and (\ref{eq:selfrs}),
the boson retarded self-energy
can be written as a single integral,
\begin{equation}
 \pir(\w)   = \displaystyle{ \sumspin \sumlr \intinf \!\!d\w' \,
        \Hlr(\w' - \w)\, \gg(\w')},
  \label{eq:Pir}
\end{equation}
with the kernels
\begin{equation}
  \begin{array}{lll}
 \Hlr(\w)   = \displaystyle{ {1 \over {(2\pi)^2}}\, \Gamalr(\w)\,
   \Biggl\{ \pi\flr(\w)}& &\\[10pt]
        \ \ \ \ \ \ \ \ \ \
     +\, \displaystyle{ i\, {\rm Re}\, \biggl\{\Psi\Bigl(\,\half -
            {{i\beta(\w - \mulr)} \over {2\pi} }\Bigr)
    -\Psi\Bigl(\,\half +
   {{\beta[W - i(\elr - \mulr)]} \over {2\pi} }\Bigr)\biggr\} }& &  \\[10pt]
        \ \ \ \ \ \ \ \ \ \
        -\, \displaystyle{ i\,{{\w - \elr} \over W} \biggl[\, {\pi \over 2}
  + {\rm Im}\, \Bigl\{\Psi\Bigl(\,\half +
       {{\beta[W - i(\elr - \mulr)]} \over {2\pi} }\Bigr)\Bigr\}
         \biggr]\Biggr\}}.& &
    \label{eq:Hspin}
   \end{array}
\end{equation}
In (\ref{eq:Hspin}), $\beta$ is the inverse temperature and
$\Psi(z)$ is the Digamma function.\cite{ams}
Second, the fermion retarded self-energies can be written
as single integrals
\begin{equation}
 \sigr(\w)   = \displaystyle{ \sumlr \intinf \!\!d\w' \,
        \Klr(\w - \w')\, \dg(\w')},
  \label{eq:Sigr}
\end{equation}
with the kernels
\begin{equation}
  \begin{array}{lll}
 \Klr(\w)   = \displaystyle{ {1 \over {(2\pi)^2}}\, \Gamalr(\w)\,
   \Biggl\{ \pi\,[1 - \flr(\w)]}& & \\[10pt]
        \ \ \ \ \ \ \ \ \ \
       +\, \displaystyle{ i\, {\rm Re}\, \biggl\{\Psi\Bigl(\,\half -
            {{i\beta(\w - \mulr)} \over {2\pi} }\Bigr)
    -\Psi\Bigl(\,\half +
    {{\beta[W - i(\elr - \mulr)]} \over {2\pi} }\Bigr)\biggr\} }& & \\[10pt]
        \ \ \ \ \ \ \ \ \ \
       +\, \displaystyle{ i\,{{\w - \elr} \over W} \biggl[\, {\pi \over 2}
  - {\rm Im}\, \Bigl\{\Psi\Bigl(\,\half +
       {{\beta[W - i(\elr - \mulr)]} \over {2\pi} }\Bigr)\Bigr\}
         \biggr]\Biggr\}}.& &
    \label{eq:Kspin}
   \end{array}
\end{equation}

Since the greater Green functions and self-energies are just the
imaginary parts of the corresponding retarded functions
(\ref{eq:ggs}-\ref{eq:selfgs}), the
above equations, together with the relation (\ref{eq:grs})
between the retarded Green functions and self-energies,
form a closed set. In practice, we make an initial guess for
the greater boson Green function and then iterate the equations
to convergence. Typically the results converge within five
iterations.  We have checked the accuracy of the results by comparing
to the sum rules on the boson and fermion Green functions
\begin{eqnarray}
 &\displaystyle{\intinf \!\! d\w
       \Bigl[ - {1 \over \pi} {\rm Im}\, \dr(\w)\Bigr] = 1}
                                                \nonumber\\[10pt]
 &\displaystyle{ \intinf \!\! d\w
       \Bigl[ - {1 \over \pi} {\rm Im}\, \gr(\w)\Bigr] = 1}.
  \label{eq:sumrules}
\end{eqnarray}
These relations are always satisfied to better than 0.5\%
by the converged numerical solutions.

A separate iterative loop is required to evaluate the
lesser Green functions and self-energies. Eqs. (\ref{eq:selfs})
for the boson and fermion lesser self-energies can be rewritten as
\begin{equation}
 \pil(\w) = \displaystyle{ - {1 \over {2\pi} } \sumspin \sumlr
  \intinf\!\!d\w'
     \,\Gamalr(\w' - \w)
    \,\Bigl[1 - \flr(\w' - \w)\Bigr]
\,\gl(\w')}
  \label{eq:Pil}
\end{equation}
and
\begin{equation}
 \sigl(\w) = \displaystyle{ - {1 \over {2\pi} } \sumlr
  \intinf\!\!d\w'
    \,\Gamalr(\w - \w')
    \,\flr(\w - \w')
  \,\dl(\w')}.
  \label{eq:Sigl}
\end{equation}
Together with Eqs. (\ref{eq:Glgs}), these form a closed set of
equations for the lesser Green functions and self-energies.
Again, following an initial guess for the boson lesser Green
function, these equations are iterated to convergence.
Since the lesser Green functions have
an arbitrary overall normalization, to check the convergence
it is necessary to monitor a normalized quantity. We choose, for
simplicity, to monitor the occupation of each spin state
\begin{equation}
  \nspin = {\Zzero \over \Zone}\, \Bigl[ -{i \over {2\pi}} \intinf \!\!
    d\w\, \gl(\w) \Bigr],
  \label{eq:occ}
\end{equation}
where the normalization is provided by the ratio of partition
functions, which from Eq. (\ref{eq:Znorm}) is given by
\begin{equation}
 {\Zone \over \Zzero} = {i \over {2\pi}} \intinf \!\! d\w\, \Bigl[
   \dl(\w) - \sumspin \gl(\w) \Bigr].
  \label{eq:norm}
\end{equation}
Typically, within five iterations the occupations converge to better
than 0.01\%. However, one cannot expect the accuracy of the
results to be better than the accuracy of 0.5\% found for the retarded
Green functions, from which the lesser functions are constructed
via (\ref{eq:Glgs}). The final accuracy is verified by
the sum-rule for infinite-$U$ relating the total density of states of one
spin state to the occupancy of all the other spin states
\begin{equation}
 \intinf \!\! d\w \,\rhosw =
            1 - \sum_{\spin' \neq \sigma} \nspinprime.
  \label{eq:sumrule}
\end{equation}
This relation is always satisfied to within $0.5\%$.

The procedure has also been checked by comparing to the
equilibrium results,\cite{coxthesis} and excellent agreement is
found. This is an independent check since much
of our numerical procedure differs from that used in equilibrium.
Importantly, because there are multiple sharp features in the
nonequilibrium density of states (discussed in the following section),
we have used a self-adjusting mesh for the numerical integrations
rather than the logarithmic mesh used in the equilibrium
case.\cite{bickers,cox}
We have also found that evaluating the
Green functions on the range $[-10W,10W]$ is sufficient for numerical
accuracy. The wider range used previously\cite{bickers,cox} is
unnecessary because the kernels (\ref{eq:Hspin}) and (\ref{eq:Kspin})
contain all the effects of the long band tails in the leads.

\subsection{Results of the Non-crossing Approximation}
\label{results}

In this section, we present numerical results of the
\ncas for an Anderson impurity in and out of equilibrium.

\subsubsection{Linear Response Conductance}

The equilibrium properties, calculated by the \nca,
 can be used to predict quantitatively
the lineshape of the linear-response conductance peak.
To our knowledge, this is the first
detailed prediction of the conductance peak evolution in
the Kondo regime. From Eq. (\ref{eq:J})
the linear-response conductance is given by
\begin{equation}
\cond = 2 \pi {{e^2}\over h} \sum_{\spin}
\int_{-\infty}^{\infty} d\w\, [ -f'(w) ]
\,\Gama(\w)\, \rhosw,
 \label {eq:conductance}
\end{equation}
where $\rhosw$ is calculated at equilibrium, and $f'(w)$ is
the derivative of the equilibrium Fermi function.
In Fig. \ref{fig:fdos} we plot
the equilibrium density of states (\ref{eq:dostwo})
of an Anderson impurity with two degenerate spin states, for
one value of chemical potential ($\mu-\ezero=2\Gamma$).
There is a sharp Kondo peak at the chemical potential.
Its amplitude increases with decreasing temperature
down to the Kondo temperature,\cite{bickers}
$\tk \simeq W (\Gamma/2\pi(\mu - \ezero))^{1/2}
\exp[-\pi(\mu-\ezero)/\Gamma]$,  where
it saturates.
In Fig. \ref{fig:fpeaks}(a), the resulting linear-response
 conductance obtained from (\ref{eq:conductance})
is plotted as the chemical potential is swept
through the bare-level energy at three different temperatures.
Several features are noteworthy. First,
as the temperature is initially lowered, the width of the conductance
peak decreases proportional to $\kbt$, because the peak lineshape is
determined by the derivative of the Fermi function
(\ref{eq:conductance}). As the temperature is lowered below $\Gamma$,
the peak width is expected to be dominated by $\Gamma$ and saturate.
Here,
however, for $\kbt \lsim 0.075 \Gamma$, the conductance peak begins to
broaden again. This broadening is entirely due
to the appearance of the Kondo peak in the density of states, and is
therefore a signature of the Kondo effect. Second, as temperature
is lowered, the peak amplitude increases and finally saturates.
The saturated open channel conductance, $\sigma = 2e^2/h$, is only
achieved for a dot symmetrically coupling to its leads; otherwise, the
conductance is reduced by the asymmetry factor
$4\Gamazl\Gamazr/(\Gamazl + \Gamazr)$. Third, the peak maximum
shifts to higher chemical potential\cite{alt} and the tails become power
law (roughly Lorentzian) rather than exponential as at higher
temperatures. It is interesting to note that the conductance peak
remains nearly symmetric with decreasing temperature despite the very
asymmetric behavior of the density of states, which has a Kondo
peak only for $\mu > \ezero$.
For comparison, the total occupancy of the site is plotted as a function
of chemical potential in
Fig. \ref{fig:fpeaks}(b). Unlike the conductance,
the total occupancy is not sensitive to the behavior
near the Fermi surface, and therefore does not show any
obvious signature of the Kondo effect.

The temperature dependences of the main features
of the conductance peak are plotted in Fig. \ref{fig:ftemp}.
Over a broad range of temperatures
the peak width, amplitude, and position increase roughly logarithmically
with temperature. This reflects the logarithmic scaling
of interactions which is the well known signature of the Kondo effect
in perturbation theory.\cite{kondo} For $\kbt \lsim 0.005 \Gamma$,
the peak amplitude saturates while the peak position and width continue
to increase. Note that the non-crossing approximation is known to
overestimate the Kondo peak amplitude somewhat for chemical potentials
within a few $\Gamma$ of the bare-level energies.\cite{friedel}
The true magnitude of the conductance peak for a symmetric structure
is therefore
not expected to approach the maximum value, $2e^2/h$, until
temperatures
below those shown in Fig. \ref{fig:fpeaks}.
For example, taking the saturated low temperature value for
$\nspin$ from Fig. \ref{fig:fpeaks}(b), which is known to be
reliable,\cite{cox} and using
Langreth's exact relation for zero temperature\cite{langreth}
\begin{equation}
\cond = 2{{e^2}\over h} \sin^2(\pi \nspin),
 \label {eq:lsumrule}
\end{equation}
one obtains a conductance $\sigma = 1.63 e^2/h$ at the peak of
the lowest temperature curve in Fig. \ref{fig:fpeaks}(a).
The observed peak value of $\sigma = 1.90 e^2/h$
is therefore 15\%
higher than the expected zero-temperature value, and must be
an overestimate. However, the tendency of the \ncas to
overestimate the conductance vanishes as $2\nspin \rightarrow 1$.

\subsubsection{Nonequilibrium}

There are
qualitatively new features in the nonequilibrium
density of states compared to equilibrium.
In Fig. \ref{fig:fdos}, the density of states (\ref{eq:dostwo})
of an Anderson impurity with two degenerate spin states
is plotted both for equilibrium and for nonequilibrium,
where the two leads have different
chemical potentials. There are striking differences between
equilibrium (solid curve) and nonequilibrium (dashed curve).
In equilibrium there is a single Kondo peak at the chemical
potential.
Out of equilibrium, the Kondo peak splits into two smaller peaks
one at each chemical potential. With decreasing temperature,
the amplitudes of these peaks {\it do not} increase to the unitarity
limit, but saturate at a much lower value. This saturation occurs
at a temperature above $\tk$, and results from dissipative processes
in which an electron is transferred from the lead with higher
chemical potential to the lead with lower chemical potential.
The \ncas includes these processes since it has contributions
from all orders in the hopping, but separating out the
relevant diagrams is not straighforward. Instead, in the next section,
we  present an analytical formula for
the dissipative lifetime obtained via perturbation theory.

A clear signature of the Kondo effect is expected in the
nonlinear current. For chemical potentials above the bare-level
energy, the linear-response conductance is dominated by the
narrow Kondo peak in the density of states. In nonlinear response,
at low temperatures, the current, $J$, is determined by an
integral of the density of states between the two chemical potentials
(\ref{eq:J}). Therefore, as soon as the chemical potential
difference exceeds the width of the Kondo peak, the differential
conductance will fall off dramatically. Moreover, the Kondo
peak will split and the split peaks decrease in amplitude with increasing
chemical potential (Fig. \ref{fig:fdos}).
The net effect is a sharp maximum peak in the differential
conductance around zero bias, for $\mu > \ezero$.\cite{selman,mwltwo}
In Fig. \ref{fig:fdc}, we have plotted the differential conductance as
a function of chemical potential difference, or equivalently
voltage bias at two temperatures.
The expected peak
is clearly resolved at a temperature $\kbt \simeq 0.05 \Gamma$.
This is substantially higher than the temperature,
$\kbt \simeq 0.025 \Gamma$,
at which the linear-response conductance peak has
broadened unambiguously (+10\%) over the minimum width.
The peak in the nonlinear differential
conductance is therefore likely to be the first signal of the
Kondo effect in transport through a quantum dot.

It is worth noting that the observability of the  Kondo peak in
the differential conductance depends only on the ratio $\kbt/\Gamma$,
not on the Kondo temperature, $\tk$. To demonstrate this,
the differential conductance is plotted at different temperatures
in Figs. \ref{fig:fdcb} and
\ref{fig:fdcc},
for level depths
differing by $\Gamma$, and, consequently, Kondo temperatures
differing by a factor of $\exp(\pi|\Delta \ezero|/\Gamma) \simeq 23$.
If the presence of a zero bias-peak depended
on the Kondo temperature one would expect the peaks to wash
out at temperatures differing by a factor of 23.
Instead, both the zero-bias peaks are clearly visible at
$\kbt = 0.05 \Gamma$ (Fig. \ref{fig:fdcb}),
but by $\kbt = 0.1 \Gamma$ (Fig. \ref{fig:fdcc}) {\it both} peaks have
washed out.
This behavior of the differential conductance reflects the
temperature dependence of the Kondo peak in the density of states.
While the final saturation of the density of states peak occurs
at temperatures below the Kondo temperature,
$\tk \sim \exp(-\pi(\mu - \ezero)/\Gamma)$, the temperature at which
the Kondo peak first appears depends only on the coupling
strength $\Gamma$. Since the peak first appears below
$\kbt \simeq 0.1 \Gamma$, the differential conductance develops
a zero-bias peak just below this temperature.

\section{Discussion}

\subsection{Theoretical Results}
\label{theory}

The most important result of this paper is a better
qualitative and quantitative understanding
of the low-temperature nonequilibrium properties of an Anderson impurity.
The nonequilibrium characteristics, in particular the transport properties,
follow from the form of the nonequilibrium density of states,
$\rhosw$.
In this section, we discuss the main features of the
density of states using both the results of the \ncas and other
methods.\cite{mwltwo,mwlone,pt}

The most obvious features in the low-temperature {\it equilibrium}
density of states for an Anderson impurity with $\mu > \ezero$
are the sharp peak at the
chemical potential and the low, broad peak around the bare-level
energy (Fig. \ref{fig:fdos}).
To understand these features it is useful to recall how, at
equilibrium,
the density of states depends on the eigenstates of the system.
At $T=0$, the density of states, $\rhosw = -(1/\pi)\, {\rm Im} \Grw$,
involves transitions from the $N$-particle ground state to all
$N+1$ or $N-1$ particle states. Since the correlated ground state
of an Anderson impurity has a finite amplitude
to have an empty site, the density of states includes a narrow
peak due to transitions
from the $N$-particle ground state to the ground state with one
more or one less electron.  By definition the ground-state energies
differ by the chemical potential, so this Kondo peak in the
density of states occurs
at the chemical potential. The weight of the Kondo
peak is small, however, since the probability that the site is
unoccupied in the ground state is much less than one.
The remaining weight, associated with transitions to excited states,
forms the low, broad peak around the bare-level energy, $\ezero$.
For finite interaction energy, $U$, there is an additional
broad feature in the density of states near $\ezero + U$;
this feature does not appear in Fig. \ref{fig:fdos} because of the
limit $U \rightarrow \infty$.

Out of equilibrium, there is no true ground state of the system,
but quantum fluctuations still produce a finite probability
of an empty site. As at equilibrium, these fluctuations involve
electrons hopping between the site and states in the leads near each
chemical potential.\cite{mwltwo,mwlone} The $N \rightarrow N+1$, and
$N \rightarrow N-1$ transitions which
determine the nonequilibrium density of states therefore include
some excitations which change the system only by
adding an electron or hole near one of the chemical potentials.
These low-energy transitions produce the Kondo peaks near each
chemical potential in the nonequilibrium density of states.

Unlike equilibrium, the configurations of the system out of equilibrium
are not true eigenstates, but have a finite lifetime, $\tau$.
The energies of transitions are therefore broadened by $\hbar/\tau$,
and all features in the density of states are broadened
an equivalent amount. This is the origin of the suppression of the
Kondo peaks out of equilibrium (Fig. \ref{fig:fdos}).
The finite lifetime is due to real processes
in which an electron is transferred from the higher chemical
potential lead to the lower chemical potential one. An estimate
of this lifetime can be obtained from straightforward perturbation
theory in the coupling strength. One assumes that the site is
initially occupied by an electron of spin $\sigma$ and calculates
the decay rate using the Golden Rule.\cite{schiff} The only
complication is that, at $T=0$, the lowest order, energy-conserving
process involves two separate tunneling events (the site electron hops
out, and another electron hops in), and
therefore occurs at $O(V^4)$. Allowing for two, possibly nondegenerate
spin states, we find
\begin{equation}
\begin{array}{ll}
\displaystyle{   {1\over{\tau_{\spin}}}  =
 {1\over{\hbar}} \sum_{{\sss A=L,R}}} \Gamma^A_{\spin}(\espin)\,& \!\!\!\!
\left[1-f_A(\espin)\right]\, + \,
   \displaystyle{{{1\over4\pi\hbar}} \,
\sum_{{A,B=L,R}\atop{\sigma'}}} \intinf\!\! d\e   \,
 \left[{1 \over {(\espin-\e + i\eta)^2}}
  + {1 \over {(\espin-\e - i\eta)^2}} \right] \, \nonumber\\
& \times \left[ \Gamma^A_{\spin}(\e)\,
\Gamma^B_{\spin'}(\e-\espin+\e_{\spin'})
 \left[1-f_A(\e)\right] f_B(\e-\espin+\e_{\spin'}) \right].
\label{eq:lifetime1}
\end{array}
\end{equation}
For a deep level, $\ezero < \mul,\mur$, so at zero temperature and for
constant $\Gamma$, Eq. (\ref{eq:lifetime1}) reduces to
\begin{equation}
{1\over{\tau_{\spin}}}  =
   \displaystyle{{{1\over2\pi\hbar}}  \sum_{{A,B=L,R}\atop{\sigma'}}}
\Gamma^A_{\spin}\,\Gamma^B_{\spin'}\,\theta(\mu_B-\mu_A+\espin-\e_{\spin'})
{{\mu_B-\mu_A+\espin-\e_{\spin'}}\over{(\mu_A-\e_{\spin})
(\mu_B-\e_{\spin'})}},
\label{eq:lifetime2}
\end{equation}
which explicitly shows that the lifetime is only nonzero for finite bias,
or finite level splitting. The results of the \ncas are consistent
with a broadening of Kondo peaks by the inverse of the
nonequilibrium lifetime $\hbar/\tau_{\spin}$. In effect,
$\hbar/\tau_{\spin}$ is an new cutoff energy for the logarithmic
scaling of interactions in the Kondo problem.\cite{kondo}

\subsection{Relation to Experiment}
\label{experiment}

Since a quantum dot weakly coupled to its leads {\it is} an
Anderson impurity, the results of the previous sections
have practical significance.
Specifically, at sufficiently low temperatures, transport through
a quantum dot will be dominated by the Kondo effect. We discuss
the practical requirements for the Kondo effect to be observed in
quantum dots, and suggest possible experiments.

There are two general classes of transport experiments to study
the Kondo effect in quantum dots: linear response and nonlinear
response. While detection of the Kondo peak in the density of states
is possible in the linear-response conductance, the nonlinear
conductance offers a clearer signature and one that persists to
higher temperatures. Figs. \ref{fig:fpeaks} and \ref{fig:ftemp}
indicate the appearance of the Kondo effect in linear response.
The sweep of chemical potential indicated in the figures can
be accomplished by sweeping the voltage of a separate gate which couples
capacitively to the dot.\cite{meirav} In fact, conductance peaks
with Lorentzian tails have already been observed in transport
through a quantum dot by this method.\cite{ethan} The long
tails of the peaks imply that coherent transport of electrons
is taking place. However, so far no broadening of the conductance peaks
at zero magnetic field
is observed down to $T \simeq 50 {\rm mK}$.
This is consistent with the prediction of the \ncas that
noticeable broadening (+10\%) occurs by $\kbt \simeq 0.025 \Gamma$,
since the largest resonance width
in the experiment is $\Gamma \sim 40 \mu{\rm eV}$ for which 10\%
broadening is not reached until $T \simeq 10 {\rm mK}$.

The appearance of the Kondo effect in nonlinear response is
shown in Figs. \ref{fig:fdc}, \ref{fig:fdcb}, and \ref{fig:fdcc}.
The sharp drop of the differential
conductance around zero applied bias reflects the sharpness of
the Kondo peak in the density of states.\cite{selman,mwltwo}
Furthermore, the peak in the
differential conductance
persists to $\kbt \simeq 0.05 \Gamma$ and therefore should be
observable in existing quantum dots up to $T \simeq 20 {\rm mK}$.
The magnitude of this zero-bias peak is optimized by performing the
differential conductance measurement at the half-maximum
point of the linear-response conductance peak (as we have done
in Fig. \ref{fig:fdc}).

An additional, striking signature of the Kondo effect in nonlinear
response is the evolution of the peak in the differential
conductance with magnetic field. From perturbation theory,\cite{pt}
and from an equations-of-motion approach,\cite{mwltwo,mwlone} it
can be shown that a finite magnetic field shifts the Kondo
peaks in the nonequilibrium densities of states by the Zeeman energy,
and consequently splits the peak in the differential conductance
by twice the Zeeman splitting of the levels.

Unfortunately, the behavior of the differential conductance
in a finite magnetic field is beyond the scope of the \nca.
Specifically, when the level degeneracy is broken, the \ncas produces,
in addition to the peaks found by the other methods,
spurious peaks in the density of states. These peaks
are due to a false interaction of each level with itself,
brought on by the neglect of vertex corrections.
While self-interaction effects are unimportant in the large-$N$ limit, for
finite $N$ the corrections can be significant.
An extreme example is the non-interacting case, $N=1$,
where the \ncas incorrectly predicts a Kondo peak in the density of states.
Because of this false self-interaction, the \ncas produces
additional Kondo peaks at the chemical potentials for non-degenerate levels,
and therefore is unreliable for transport properties in a magnetic
field.  Interestingly, the \ncas continues to produce reliable results
in a magnetic field for thermodynamic quantities
({\it e.g.} magnetization\cite{cox})
which depend on the entire density of states and not just on the behavior
near the Fermi surface.

\section{Conclusion}

In summary, we have analyzed the low-temperature, nonequilibrium properties
of an Anderson impurity in the limit of infinite on-site interaction.
The model corresponds to a quantum dot,
weakly coupled to two leads with different chemical potentials.
The Kondo effect, which dominates transport
through the impurity, is modified by two new energies present in
nonequilibrium: the chemical potential difference $\Delta \mu$, and
the inverse of the dissipative lifetime $\hbar/\tau_{\spin}$
(\ref{eq:lifetime1}). These energies are apparent in the
nonequilibrium density
of states, which we obtain via the non-crossing approximation.
The chemical potential difference appears in the density of states
via the splitting of the Kondo peak into two peaks, one at each
chemical potential. The amplitudes of these peaks are suppressed
by dissipative processes in which an electron is
transferred from the higher chemical potential lead to the lower
chemical potential one.

Experimentally, we predict that the Kondo effect can be observed
in transport through a quantum dot by either linear or nonlinear
measurements. The emergence of the Kondo peak in the density
of states at low temperatures will cause the linear-response
conductance peaks vs. gate voltage to broaden, shift, and increase
in amplitude roughly logarithmically with decreasing temperature.
 For a symmetric structure the conductance amplitude
will saturate at $2e^2/h$, the conductance of an open
channel.\cite{glazman,ng}
The clearest feature of the Kondo effect in linear response, however,
is the broadening of the conductance peak, which is predicted
to reach +10\% below
$\kbt \simeq 0.025 \Gamma$, where $\Gamma$ is the
total coupling strength to the leads.
In nonlinear response, the Kondo peak will produce a peak in
the differential conductance around zero bias.\cite{selman}
Since this nonlinear peak remains clearly defined for temperatures up
to $\kbt \simeq 0.05 \Gamma$, we believe it will be the most accessible
signature of the Kondo effect in quantum dots.

We thank Dan Cox, Roger Lake, and Patrick A. Lee for valuable
discussions. Work at U.C.S.B. was supported by NSF Grant No.
NSF-DMR90-01502 and by the NSF Science and Technology Center
for Quantized Electronic Structures, Grant No. DMR 91-20007.

\newpage
\appendix{}
\section{}

In this appendix, we show that the \ncas is current conserving.
Specifically, the current through the Anderson impurity can be expressed
either as a current flowing from the left lead into the site
or as a current flowing from the site into the right lead.
Within the \nca, these two expressions are equivalent.

The full expression for the current through the left(right)
tunneling barrier, with no restriction on the relative couplings
to the leads,\cite{mwone} is
\begin{equation}
   J_{\sss L(R)} = \displaystyle{ +(\!-\!) \,{{ie} \over h}\,
     \sum_{\spin} \intinf\!\! d\w
       \,\Gamalr(\w)\, \biggl\{ \Bigl[1 - \flr(\w)\Bigr]\, \Gl(\w) + \flr(\w)
             \,\Gg(\w)
           \biggr\}    },
 \label {eq:currentlr}
\end{equation}
where $\Glg(\w)$ are Fourier transforms of the
physical-electron Green functions defined in (\ref{eq:Gless}).
In the \nca, these Green functions factorize into boson
and fermion parts,
\begin{equation}
 \Glg(\w)   \stackrel{\rm {\sss NCA}}{=}
 \displaystyle{ {i \over {2\pi}}\, {\Zzero \over \Zone}\,
\intinf \!\!d\w' \,\dgl(\w')\, \glg(\w + \w')}.
 \label {eq:Gphyslg}
\end{equation}
Since this factorization is an approximation, one can ask whether
the two expressions for the current (\ref{eq:currentlr}) remain
identical.

To show that current is conserved in the \nca, we examine the
difference, $\jl - \jr$, between the currents flowing through the
two tunneling barriers,
\begin{equation}
\begin{array}{ll}
\jl- \jr = \displaystyle{ -{e \over {2\pi h}}\, {\Zzero \over \Zone}\,
 \sumspin \sumlr \intinf\!\! d\w \intinf\!\! d\w' \,\Gamalr(\w)   }
                                  \nonumber\\
 \ \ \ \ \
    \times \biggl\{\Bigl[1 - \flr(\w)\Bigr]\, \dg(\w') \, \gl(\w + \w')
           +        \flr(\w) \, \dl(\w') \, \gg(\w + \w') \biggr\}.
\label{eq:jljr1}
\end{array}
\end{equation}
This expression is simplified by recognizing that the
$\w$-integration produces factors of the boson self-energies,
\begin{equation}
\begin{array}{ll}
 \pig(\w') = &\displaystyle{  {1 \over {2\pi} } \sumspin \sumlr
  \intinf\!\!d\w
     \,\Gamalr(\w)
    \,\flr(\w)
\,\gg(\w + \w')}\nonumber\\
 \pil(\w') = &\displaystyle{ - {1 \over {2\pi} } \sumspin \sumlr
  \intinf\!\!d\w
     \,\Gamalr(\w)
    \,\Bigl[1 - \flr(\w)\Bigr]
\,\gl(\w + \w')}.
  \label{eq:Pigl}
\end{array}
\end{equation}
The difference can therefore be written as
\begin{equation}
\jl - \jr = \displaystyle{  {1 \over {2\pi} }\, {\Zzero \over \Zone} }\,
  \intinf\!\!d\w'\, \Bigl[ \dg(\w')\, \pil(\w') - \dl(\w')\, \pig(\w') \Bigr].
  \label{eq:jljr2}
\end{equation}
The integrand vanishes because of the relation between the
boson Green functions and self-energies  (\ref{eq:Glgs}),
\begin{equation}
 \dlg(\w) = \dr(\w)\, \pilg(\w)\, \da(\w).
  \label{eq:dgreat}
\end{equation}
The \ncas therefore explicitly conserves current,
\begin{equation}
 \jl = \jr.
  \label{eq:jljr3}
\end{equation}

\newpage
\begin{figure}[h]
\bigskip
\caption{
\baselineskip 24pt
\hsize 15 truecm
\hoffset -1.0 truecm
Schematic band diagram of a quantum dot coupled via tunneling
barriers to two leads with different chemical potentials. At zero
magnetic field, the energy level $\ezero$ on the quantum dot
will be spin degenerate, and a large Coulomb interaction energy, $U$,
will prevent double occupancy. }
\label{fig:schematic}
\end{figure}

\begin{figure}[h]
\caption{
\baselineskip 24pt
\hsize 15 truecm
\hoffset -1.0 truecm
Diagrammatic expansion for (a) the slave boson and (b) the fermion
propagators. The coupling between site and leads is treated
as the perturbation, so each vertex corresponds to a tunneling event.}
\label{fig:pt}
\end{figure}

\begin{figure}[h]
\caption{
\baselineskip 24pt
\hsize 15 truecm
\hoffset -1.0 truecm
Real-time contour for
nonequilibrium Green functions in the Keldysh formalism.\ \ }
\label{fig:contour}
\end{figure}

\begin{figure}[h]
\caption{
\baselineskip 24pt
\hsize 15 truecm
\hoffset -1.0 truecm
Diagrammatic representation of the non-crossing approximation.
(a) Dyson's equation for the boson propagator includes the fermion
propagators in the self-energy, and (b) Dyson's equation for each
fermion propagator includes the boson propagator in the self-energy. }
\label{fig:nca}
\end{figure}

\begin{figure}[h]
\caption{
\baselineskip 24pt
\hsize 15 truecm
\hoffset -1.0 truecm
Equilibrium and nonequilibrium density of states  $\rhosw$
for an Anderson impurity symmetrically
coupled to two leads of Lorentzian bandwidth $2W$ and
chemical potentials $\mul$ and $\mur$.
The impurity has two degenerate spin states
at energy $\ezero = 0$,
and an on-site interaction $U\rightarrow\infty$.
With all energies in units of the total coupling to the leads, $\Gamma$,
the band
half-width is $W=100$ and the temperature is $T=0.005$.
At equilibrium (solid curve) there is a single Kondo peak in the
density of states at the chemical potential $\mul = \mur = 2$
(see inset).
Out of equilibrium (dashed curve),
the peak splits into two suppressed peaks,
one at each chemical potential, $\mul = 2.4$ and $\mur = 2$.}
\label{fig:fdos}
\end{figure}

\begin{figure}[h]
\caption{
\baselineskip 24pt
\hsize 15 truecm
\hoffset -1.0 truecm
(a) Linear-response conductance $\cond$ through an Anderson
impurity for three different temperatures as a function of
chemical potential. The impurity has two degenerate spin
states at $\ezero = 0$.
The conductance peak
first narrows then broadens with decreasing temperature.  (b)
Total site occupancy, $\nup + \ndown$, as a function of chemical
potential for the same temperatures.}
\label{fig:fpeaks}
\end{figure}

\begin{figure}[h]
\caption{
\baselineskip 24pt
\hsize 15 truecm
\hoffset -1.0 truecm
(a) Temperature dependence of linear-response conductance peak
position.
(b) Temperature dependence of conductance peak amplitude.
(c) Temperature dependence of conductance peak full-width at
half maximum. In all three panels, the
non-crossing approximation results are the data
points and the solid curve is a guide to the eye. For comparison,
the dashed curves are the exact results for non-interacting levels.}
\label{fig:ftemp}
\end{figure}

\begin{figure}[h]
\caption{
\baselineskip 24pt
\hsize 15 truecm
\hoffset -1.0 truecm
Differential conductance, $e\,dJ/d\Delta\mu$,  with $\mur=1.9$,
vs. applied bias, at two
temperatures, $\kbt = 0.005$ and $\kbt = 0.05$.
The peak in the differential conductance at zero bias
reflects the Kondo peak
in the density of states.}
\label{fig:fdc}
\end{figure}

\begin{figure}[h]
\caption{
\baselineskip 24pt
\hsize 15 truecm
\hoffset -1.0 truecm
Differential conductance, $e\,dJ/d\Delta\mu$
vs. applied bias for $\mur = 1.9$ (solid curve) and $\mur = 2.9$
(dashed curve) at $\kbt = 0.05$. Zero-bias peaks due to the Kondo
effect appear for both curves despite the very
different Kondo temperatures.}
\label{fig:fdcb}
\end{figure}

\begin{figure}[h]
\caption{
\baselineskip 24pt
\hsize 15 truecm
\hoffset -1.0 truecm
Differential conductance, $e\,dJ/d\Delta\mu$
vs. applied bias for $\mur = 1.9$ (solid curve) and $\mur = 2.9$
(dashed curve) at $\kbt = 0.1$. Both zero-bias peaks
in Fig. 9 have become shoulders.}
\label{fig:fdcc}
\end{figure}


\begin{references}
\bibitem{baregs}
 It would not be correct in (\ref{eq:selfs}) to include self-energies
 for the Green functions in the leads. Such self-energies include
 a fermion-boson loop which contributes an extra,
 unwanted
 factor of
 $\exp(-i\beta\lambda)$.
\end{references}
\end{document}